# Phát hiện đánh giá spam cho tiếng Việt

(Detecting Vietnamese Opinion Spam)

Dương Hiển Hồng Thạch[1,2,3]
hongthach.duong@gmail.com

Vũ Đại Thắng[1,2,3]
daithang.vu@gmail.com

Ngô Minh Vương[1,2,3]
vuong.cs@gmail.com

[1] Phòng Nghiên Cứu và Phát Triển, tập đoàn VNG
[2] Khoa Công Nghệ Thông Tin, Đại học Tôn Đức Thắng
[3] Khoa Khoa Học và Kỹ Thuật Máy Tính, Đại Bách Khoa TpHCM

**Abstract.** Recently, Vietnamese Natural Language Processing has been researched by experts in academic and business. However, the existing papers have been focused only on information classification or extraction from documents. Nowadays, with quickly development of the e-commerce websites, forums and social networks, the products, people, organizations or wonders are targeted of comments or reviews of the network communities. Many people often use that reviews to make their decision on something. Whereas, there are many people or organizations use the reviews to mislead readers. Therefore, it is so necessary to detect those bad behaviors in reviews. In this paper, we research this problem and propose an appropriate method for detecting Vietnamese reviews being spam or non-spam. The accuracy of our method is up to 90%.

*Abstract -* *Trong những năm gần đây, xử lý ngôn ngữ văn bản tiếng Việt đã thu hút đối với nhiều chuyên gia trong giới học thuật cũng như trong công nghiệp. Các công trình trước đây chủ yếu tập trung vào việc phân loại hay rút trích thông tin từ văn bản. Chúng không phải là nhận diện tình cảm được thể hiện trong văn bản. Ngày nay với sự phát triển của các website thương mại điện tử, các diễn đàn, mạng xã hội tiếng Việt nên có nhiều đối tượng như sản phẩm, con người, tổ chức, thắng cảnh được người dùng bình luận và đánh giá trên các website này. Nhiều người thường sử dụng các đánh giá này để đưa ra quyết định của họ. Trong khi đó, có nhiều cá nhân, tổ chức sử dụng các đánh giá bình luận nhằm mục đích đánh lừa người khác. Vì vậy việc phát hiện hành vi spam trong các đánh giá là điều cần thiết. Công trình của chúng tôi tập trung vào giải quyết vấn đề trên và đề xuất một phương pháp phù hợp cho việc đánh giá các bình luận tiếng Việt là spam hay không spam, với độ chính xác 90%.*

***Từ khóa:*** *classification, spam review, opinion spam*

## I. GIỚI THIỆU

Đánh giá của người dùng về sản phẩm là những nhận định của người dùng về sản phẩm ấy. Người đọc sẽ tham khảo đánh giá để quyết định có nên mua sản phẩm hay không. Hãng sản xuất tham khảo đánh giá để quyết định chiến lược kinh doanh phù hợp cho sản phẩm. Nhìn chung, một đánh giá có những thành phần sau ([6]):

- Tiêu đề: mô tả ngắn về đánh giá
- Nội dung: ý kiến chính của người dùng
- Ngày giờ: thời gian đánh giá xuất hiện
- Sản phẩm: đối tượng của đánh giá
- Người đánh giá: tác giả của đánh giá
- Điểm số: thang điểm người dùng dành cho sản phẩm

Đánh giá spam là những đánh giá chứa hành vi spam. Những hành vi này (thường vì mục đích thương mại hay mục đích cá nhân của người viết) nhằm thay đổi nhìn nhận của người đọc về sản phẩm được đánh giá. Cũng theo [6], có ba loại đánh giá spam:

1) *Những đánh giá không thực:* loại đánh giá này thường đưa ra những nhận xét tích cực quá mức hoặc đưa ra những nhận xét tiêu cực quá khắt khe. Chúng sẽ làm danh tiếng sản phẩm được nâng lên hoặc bị phá hoại. Đây là loại khó phát hiện nhất của đánh giá spam.

2) *Những đánh giá vào hãng sản xuất:* đây mặc dù là những đánh giá chứa nhận định khen chê, nhưng chúng không hướng vào sản phẩm mà lại hướng vào hãng sản xuất, cửa hàng bán sản phẩm. Đôi khi những đánh giá này cũng gây ra sự nhầm lẫn cho người đọc. Ví dụ với đánh giá sau: "Sony là laptop số một, không laptop nào qua được sony" sẽ được xem là đánh giá spam loại này.

3) *Những đánh giá vô nghĩa:* loại đánh giá này thường không chứa nhận định và không có giá trị tham khảo. Đây thường là những câu hỏi, những thông số kỹ thuật, quảng cáo,…

## II. CÁC NGHIÊN CỨU LIÊN QUAN

Trong phần này, chúng tôi sẽ trình bày những nghiên cứu liên quan đến lĩnh vực phát hiện đánh giá spam.

*Khai phá quan điểm và nhận định của đánh giá:* trong lĩnh vực này, đã có những nghiên cứu cố gắng rút ra và tập hợp các đánh giá tích cực và tiêu cực từ các đánh giá sản phẩm [5, 14]. Những nghiên cứu đó tập trung chủ yếu vào nội dung của đánh giá và rất hữu dụng trong việc xác định các quan điểm trái chiều trong đánh giá. Tuy nhiên những nghiên cứu này chưa phát hiện ra được đánh giá spam, trừ khi có thêm những đặc trưng khác.

*Phát hiện ba loại đánh giá spam:* trong nghiên cứu [6], tác giả định nghĩa ba loại đánh giá spam như đã nêu trong phần giới thiệu. Tác giả tạo ra một tập đặc trưng kết hợp các thuộc tính của đánh giá, người đánh giá và sản phẩm. Sau đó tập đặc trưng này được tác giả sử dụng kết hợp với các kỹ thuật phân loại để gán nhãn cho từng đánh giá. Những đánh giá vào hãng sản xuất và những đánh giá vô nghĩa sẽ sử dụng một tập huấn luyện được gán nhãn bằng tay, trong khi việc phát hiện đánh giá không thực khó có thể gán nhãn bằng tay nên tác giả sẽ dùng những đánh giá trùng lặp để huấn luyện.

*Phát hiện spam dựa trên hành vi bất thường của người đánh giá:* có hai phương pháp phổ biến được giới thiệu là phương pháp *chấm điểm hành vi* đã được giới thiệu trong [11] và phương pháp *sử dụng luật không kỳ vọng* được nghiên cứu trong [7]. Phương pháp thứ nhất định nghĩa ra bốn hành vi bất thường là *nhắm mục tiêu vào sản phẩm, nhắm mục tiêu vào nhóm sản phẩm, chấm điểm lệch* và *cho điểm lệch sớm,* sau đó tạo ra một hàm đánh giá để chấm điểm từng người dùng dựa trên bốn hành vi này. Phương pháp thứ hai sử dụng luật không kỳ vọng để đánh giá người dùng. Tập dữ liệu dùng cho phương pháp này được mô tả bởi một tập thuộc tính dữ liệu: A = {$A_1$, $A_2$,…, $A_n$} và thuộc tính phân lớp C = {$C_1$, $C_2$,…, $C_n$}. Từ hai tập thuộc tính này sẽ tạo ra được một tập luật, mỗi luật có dạng X→$c_i$ trong đó X là một tập thuộc tính của A và $c_i$ là một thuộc tính thuộc C. Mỗi luật sẽ có một xác suất có điều kiện P($c_i$ | X) gọi là độ tin cậy và một xác suất hợp P(X, $c_i$) để X và $c_i$ xảy ra đồng thời gọi là độ hỗ trợ. Tác giả sử dụng bốn luật không kỳ vọng khác nhau để đánh giá dựa trên bốn định nghĩa độ không kỳ vọng cho *độ tin cậy, độ hỗ trợ, sự phân bố của thuộc tính*, và cho *thuộc tính*.

*Phát hiện spam theo nhóm:* phương pháp này được giới thiệu trong [1], tập trung nghiên cứu về một nhóm người cùng viết những đánh giá sai sự thực để đạt được mục đích cá nhân của họ. Việc phát hiện các nhóm spam này được thực hiện qua ba bước. Đầu tiên, từ các bộ dữ liệu gồm một sản phẩm và toàn bộ ID của người đã đánh giá sản phẩm đó, ta tìm ra những nhóm ứng cử viên có khả năng là spam. Tiếp theo, kiểm tra lại các nhóm ứng viên này bằng một vài tiêu chí được đề xuất. Cuối cùng, xếp hạng các nhóm ứng viên theo các tiêu chí ở bước 2.

*Phát hiện đánh giá không thực:* phương pháp này đã được giới thiệu trong [13]. Đầu tiên, những đánh giá cần phân loại sẽ được thực hiện thủ công do con người. Sau đó, tác giả sử dụng ba tập đặc trưng khác nhau: N-Gram, LIWC và Part-Of-Speech (POS). Dựa vào các tập đặc trưng này, các đánh giá sẽ được dùng để huấn luyện các mô hình phân loại. Kết quả của phương pháp này cho thấy hiệu suất phân loại của máy tốt hơn hẳn khả năng phân loại của con người, nghĩa là con người rất khó phát hiện những đánh giá không thực này.

*Dự đoán đánh giá hữu ích:* nghiên cứu này được giới thiệu trong [2]. Tác giả nhận thấy có mối liên hệ lớn giữa tỷ lệ các đánh giá được bình chọn là hữu ích với độ lệch điểm số của đánh giá đó so với điểm trung bình. Mối liên hệ đó chỉ ra rằng, các đánh giá hữu ích thường đi cùng với một điểm số trung bình. Tuy nhiên nghiên cứu này lại không đưa ra được dẫn chứng chứng tỏ mối liên kết giữa spam và các đánh giá hữu ích.

*Nghiên cứu về đánh giá cho ngôn ngữ của các nước châu Á:* ngoài những nghiên cứu đã được thực hiện cho tiếng Anh ở trên, việc nghiên cứu đánh giá của người dùng cũng đã được thực hiện cho một vài ngôn ngữ khác như tiếng Trung Quốc, Nhật hay Việt Nam. Ví dụ như trong [18] là nghiên cứu *rút trích nhận định và đặc trưng sản phẩm* cho tiếng Trung Quốc. Tác giả đề xuất ba bước để rút trích đặc trưng sản phẩm kết hợp với hướng nhận định của người dùng: (i) xác định các đặc trưng sản phẩm, (ii) xác định các nhận định liên quan đến sản phẩm, và (iii) xác định hướng tình cảm cho các sản phẩm đó. Rút trích nhận định cho tiếng Nhật được trình bày trong [10], với ý tưởng chính là rút ra bốn bộ dữ liệu với mỗi đánh giá: {Sản phẩm, Thuộc tính, Giá trị, Lượng giá}. Sau đó, tác giả sử dụng các phương pháp học máy dùng cho phân giải trùng lặp để giải quyết vấn đề. Trong nghiên cứu [9], tác giả đã đề xuất cách để rút trích nhận định trong đánh giá bằng cách phát hiện những từ chứa nhận xét của người dùng, từ đó nhận dạng những câu chứa các từ đó rồi sau đó thống kê lại để tìm ra ý kiến cuối cùng của người đánh giá dành cho sản phẩm. Việc tìm câu chứa nhận định được tác giả dựa trên hướng tiếp cận luật.

### III. CƠ SỞ LÝ THUYẾT

*A. Học máy vector hỗ trợ (Support Vector Machine, SVM )*

SVM là phương pháp phân loại sẽ nhận một tập dữ liệu đầu vào và xác định từng phần tử thuộc về lớp nào trong hai lớp cho sẵn dựa trên tổ hợp tuyến tính của các giá trị đặc trưng. Cho trước một tập huấn luyện được biểu diễn trong không gian vector, SVM sẽ tìm ra một siêu mặt phẳng tối ưu nhất có thể chia các điểm trên không gian này thành hai lớp riêng biệt tương ứng. Ví dụ trong Hình 1, ta có thể thấy sẽ có rất nhiều mặt phẳng có khả năng phân tách dữ liệu thành hai lớp (H1, H2 và H3) và cũng có những mặt phẳng không thể phân tách (H4). SVM sẽ tìm trong các mặt phẳng phân tách đó và chọn ra mặt phẳng tối ưu nhất. Với phương trình mặt phẳng trong không gian có dạng:

$$w \bullet x - b = 0 \qquad (1)$$

Trong đó, w là vector pháp tuyến của mặt phẳng, x là vector các giá trị đặc trưng, b là hệ số tự do. Mục tiêu của SVM là tìm w và b sao cho đạt được cực đại khoảng cách từ mặt phẳng phân tách này đến các điểm gần nhất của mỗi lớp ([17]).

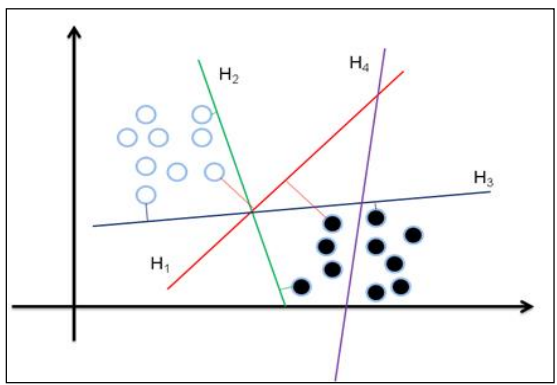

Hình 1. Một số mặt siêu phẳng phân tách SVM

*B. Hồi quy Logistic (Logistic Regression - LR)*

Đây là kỹ thuật dùng để phân tích dữ liệu với đầu vào là một hay nhiều biến (đặc trưng, thuộc tính) độc lập và đầu ra là một biến phụ thuộc (biến phụ thuộc chỉ mang hai giá trị). Công thức của LR như sau ([17]):

$$p(X) = \frac{1}{1+e^{\beta_0+\beta_1 x_1+\ldots+\beta_n x_n}} \quad (2)$$

Trong đó, $p(X)$ là xác suất dữ liệu $X$ thuộc lớp quan tâm. Vì vậy xác suất để dữ liệu $X$ thuộc lớp không quan tâm là $1 - p(X)$. Các hệ số $\beta$ được ước tính từ dữ liệu. Các giá trị $x$ là trọng số của vector dữ liệu đầu vào.

Hình 2 là đồ thị của hàm hồi quy logistic. Đầu vào của dữ liệu là không có giới hạn, và đầu ra là giá trị nằm trong khoảng (0,1). Do đó LR được gọi là bộ phân loại xác suất.

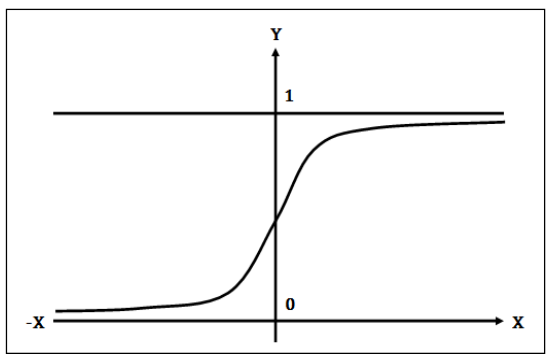

Hình 2. Đồ thị của hàm hồi quy logistic

*C. Phương pháp Naïve Bayes (NB)*

Như tên gọi của giải thuật, ý tưởng chủ yếu ở đây là việc sử dụng định luật Bayes:

$$P(Y=c_i \mid X) = \frac{P(X \mid Y=c_i) \times P(Y=c_i)}{P(X)} \quad (3)$$

Giải thuật sẽ tính tất cả xác suất với dữ liệu đầu vào là vector $X$ và đầu ra là $c_i$. Sau đó giá trị $c_i$ làm cực đại xác suất trên sẽ là loại của dữ liệu ([8]).

## IV. SO SÁNH GIỮA ĐÁNH GIÁ TIẾNG ANH VÀ ĐÁNH GIÁ TIẾNG VIỆT

Các nghiên cứu về lĩnh vực phát hiện đánh giá spam này đã được thực hiện nhiều trên tiếng Anh với dữ liệu được lấy từ các trong web bán hàng, giới thiệu sản phẩm trực tuyến. Tuy nhiên, giữa các trang web tiếng Anh được khảo sát trong các nghiên cứu đó với các trang web giới thiệu sản phẩm tiếng Việt có nhiều điểm khác nhau cả về hệ thống đánh giá lẫn nội dung đánh giá.

Những điểm khác nhau nêu trên khiến việc phát hiện đánh giá spam trên tiếng Việt gặp nhiều khó khăn. Ví dụ như trong Bảng 1, khác biệt (I) và (IV) khiến việc phát hiện spam dựa trên hành vi người dùng không thể thực hiện. Khác biệt (II) và (V) gây ra khó khăn trong việc phát hiện đánh giá không thật. Còn khác biệt (III) dẫn đến khó phát hiện đánh giá spam dựa trên độ lệch điểm số. Tóm lại, với đánh giá tiếng Việt, hướng tiếp cận chủ yếu chỉ là nội dung của đánh giá, còn việc tiếp cận theo thời gian đánh giá được đưa lên, điểm số sản phẩm hay hành vi người dùng là rất khó khả thi.

Bảng 1. So sánh đánh giá tiếng Anh và tiếng Việt

| Điểm khác nhau | Trang web tiếng Anh | Trang web tiếng Việt |
|---|---|---|
| Chức năng cho phép người dùng phản hồi đánh giá có ích hay không (I) | Có và thường được sử dụng | Hầu như không có, nếu có cũng sơ sài và ít được sử dụng |
| Các sản phẩm chưa bắt đầu được phân phối (II) | Sản phẩm không được đưa lên giới thiệu khi chưa bắt đầu bán | Sản phẩm được đưa lên giới thiệu từ rất sớm trước khi được bán chính thức |
| Người dùng cho điểm sản phẩm (III) | Người dùng cho điểm rất nghiêm túc (ngoại trừ những người spam) | Người dùng ít quan tâm cho điểm, đa số là điểm mặc định từ trang web |
| Quản lý thông tin người dùng (IV) | Phải đăng ký trước khi đưa ra đánh giá, thông tin người dùng được lưu giữ chặt chẽ | Không cần đăng ký, chỉ cần nhập một tên ngẫu nhiên cũng có thể đánh giá sản phẩm. |
| Những đánh giá không thật (V) | Chiếm tỷ lệ lớn | Tỷ lệ thấp, đa số là những đánh giá vô nghĩa |

## V. PHƯƠNG PHÁP ĐỀ XUẤT

Bài toán nhận dạng đánh giá spam được mô tả như bài toán phân loại và chúng sẽ được giải quyết bằng học máy.

## A. Chuẩn bị tập dữ liệu

Chúng tôi đã sử dụng tập dữ liệu là những đánh giá về sản phẩm từ những trang web bán hàng trực tuyến của Việt Nam. Một số trang web tiêu biểu có thể kể đến như: www.thegioididong.com, www.voz.vn, www.mainguyen.vn, www.vatgia.com, www.aha.vn... Tập dữ liệu chúng tôi sử dụng gồm 2.000 đánh giá. Tuy nhiên, những đánh giá từ các trang web thường được viết dưới dạng ngôn ngữ không phải là tiếng Việt chuẩn ("wá", "hok", "ko"). Chính vì vậy, chúng tôi đã thực hiện bước tiền xử lý là chuẩn hóa lại những từ như trên ("wá" → "quá", "hok, ko" → "không").

## B. Tạo tập đặc trưng

Khi sử dụng học máy để phân loại dữ liệu, việc xây dựng tập đặc trưng là nhiệm vụ quan trọng nhất. Vì những hạn chế của dữ liệu đánh giá tiếng Việt, chúng tôi đã đề xuất tập đặc trưng chỉ dựa trên nội dung các đánh giá. Tập đặc trưng chúng tôi sử dụng gồm hai loại: các từ được tách từ tập dữ liệu và các từ đại diện cho đánh giá spam (và không spam) được chúng tôi liệt kê bằng tay.

*Các từ được tách từ tập dữ liệu*: sở dĩ chúng tôi sử dụng các từ này là để quan sát tần số xuất hiện của chúng trong từng loại đánh giá spam và không spam. Từ đó dùng chúng để nhận dạng loại đánh giá. Ví dụ với câu "Điện thoại này đẹp", nó sẽ được tách thành ba từ: [Điện thoại], [này] và [đẹp]. Trong hiện thực, chúng tôi đã sử dụng lại phần mềm JvnTextPro. Đây là phần mềm tách từ cho tiếng Việt với kết quả đạt được khá tốt ([16]). Từ tập huấn luyện gồm 1.000 câu đánh giá với khoảng 2.790 từ sẽ được tách ra. Các câu trong tập huấn luyện này được chúng tôi phân loại spam hay không spam bằng tay. Tuy nhiên nếu sử dụng tất cả các từ tách được, mô hình phân loại có thể không tốt trên tập kiểm thử do một số từ không có khả năng phân loại. Ví dụ như những từ xuất hiện ở cả hai lớp, hoặc những từ xuất hiện rất ít trong tập dữ liệu. Chính vì vậy, chúng tôi đã thực hiện giảm những từ không cần thiết để cải thiện nhược điểm trên. Chúng tôi tính điểm cho từng từ dùng Chi-Square hoặc Odd-Ratio, sau đó lấy $k$ từ có số điểm cao nhất (có khả năng phân loại nhất).

Bảng 2. Thống kê sự xuất hiện của đặc trưng (từ) $t$

| Lớp | Chứa $t$ | Không chứa $t$ |
|---|---|---|
| Lớp $c$ | A | C |
| Không thuộc lớp $c$ | B | D |

Trong Bảng 2, $A$ là số đánh giá chứa $t$ và thuộc lớp $c$, $B$ là số đánh giá chứa $t$ nhưng không thuộc lớp $c$, $C$ là số đánh giá không chứa $t$ nhưng thuộc lớp $c$, $D$ là số đánh giá không chứa $t$ và không thuộc lớp $c$. Ta có công thức để tính điểm theo Chi-Square như sau ([3]):

$$\chi^2(t,c) = \frac{(A+B+C+D) \times (A \times D - B \times C)^2}{(A+C) \times (A+B) \times (D+C) \times (D+B)} \quad (4)$$

Và công thức để tính điểm theo Odd-Ratio như sau ([3]):

$$OR(t,c) = \frac{A \times D}{B \times C} \quad (5)$$

*1) Các từ được liệt kê bằng tay:* chúng tôi cũng đã liệt kê những từ thường hay dùng trong đánh giá không spam và những từ hay dùng trong đánh giá spam. Nếu tỉ lệ từ chứa nhận định ví dụ như "tốt", "bắt mắt", "đẹp" xuất hiện càng nhiều thì khả năng đánh giá đó là không spam càng cao. Ngược lại, nếu từ nghi vấn ví dụ như "khi nào", "ở đâu", "bao giờ" chiếm tỉ lệ lớn thì đánh giá đó nhiều khả năng sẽ là spam. Chúng tôi đã liệt kê khoảng 170 từ chứa nhận định và khoảng 30 từ nghi vấn.

## C. Biểu diễn đánh giá dưới dạng vector

Để các giải thuật phân loại có thể sử dụng được tập huấn luyện, các đánh giá phải được chuyển về dạng vector. Số chiều của vector này bằng số đặc trưng đã tạo được ở trên.

*1) Đối với từ được tách từ tập dữ liệu:* trọng số của chúng sẽ được tính dựa vào tích TFIDF([15]). Từ xuất hiện càng nhiều trong đánh giá đầu vào và xuất hiện trong càng ít trong đánh giá của tập huấn luyện thì khả năng từ đó dùng để nhận dạng loại đánh giá càng tốt.

*2) Đối với từ được liệt kê bằng tay:* trọng số của các từ này sẽ được tính bằng tỉ lệ giữa số lần từ này xuất hiện và tổng số từ của đánh giá đang xét.

## D. Xây dựng mô hình phân loại

Chúng tôi lần lượt áp dụng ba giải thuật phân loại như đã nêu ở phần trước (được tích hợp trong phần mềm mã nguồn mở *Weka[1]*) để xây dựng mô hình phân loại. Qua đó chúng tôi sẽ so sánh kết quả nhận được từ các giải thuật khác nhau. Sau khi các mô hình đã được xây dựng, các đánh giá cần phân loại sẽ được chuyển về dạng vector và mô hình sẽ dựa vào vector này để trả về kết quả là spam hay không spam.

## VI. KẾT QUẢ THỰC NGHIỆM

Với 2.000 đánh giá trong tập dữ liệu, chúng tôi chia theo tỉ lệ 1:1 để làm tập huấn luyện và tập kiểm thử. Các dữ liệu đã được gán nhãn spam và không spam bằng tay, cũng theo tỉ lệ 1:1. Đối tượng chủ yếu trong các đánh giá đã được thu thập là những sản phẩm về thiết bị kỹ thuật số, xe máy, xe ô tô.

Chúng tôi sử dụng các độ đo P, R, F ([12]) và AUC ([4]) để so sánh các kết quả thu được:

- Độ P: tỉ lệ giữa số đánh giá được phân loại thật sự đúng (dựa vào loại của đánh giá đã được gán trước đó) và số đánh giá được máy trả về là đúng.
- Độ R: tỉ lệ giữa số đánh giá được phân loại thật sự đúng và số đánh giá đúng cần được phân loại.
- Độ F: cân bằng giữa độ P và độ R.
- Độ AUC: xác suất bộ phân loại xếp hạng một phần tử ngẫu nhiên của lớp quan tâm cao hơn một phần tử ngẫu nhiên của lớp không quan tâm.

---

[1] www. www.cs.waikato.ac.nz/ml/weka

Chúng tôi sẽ sử dụng những ký hiệu sau trong phần kết quả thực nghiệm:

- (1): Chỉ lấy k từ tách được (trong bài báo này, chúng tôi sử dụng k = 500).
- (2): Từ được chọn theo phương pháp Chi-Square.
- (3): Từ được chọn theo phương pháp Odd-ratio.
- (4): Có sử dụng các từ được liệt kê bằng tay.

Trong Hình 3, Hình 4 và Hình 5 lần lượt là kết quả phân loại của các giải thuật SVM, NB và LR. Kết quả phân loại thay đổi khi số lượng đặc trưng (tách từ tập dữ liệu) thay đổi. Như đã nói ở trên, những từ xuất hiện đồng đều ở cả hai lớp hay những từ ít xuất hiện sẽ không có khả năng phân loại. Từ kết quả phân loại của ba giải thuật, chúng ta thấy với giải thuật SVM cùng 500 đặc trưng có thì kết quả đạt được là tốt nhất.

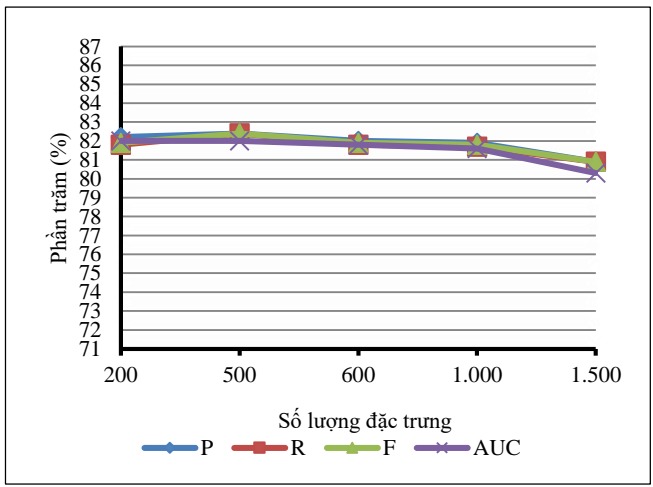

Hình 3. Kết quả phân loại của giải thuật SVM

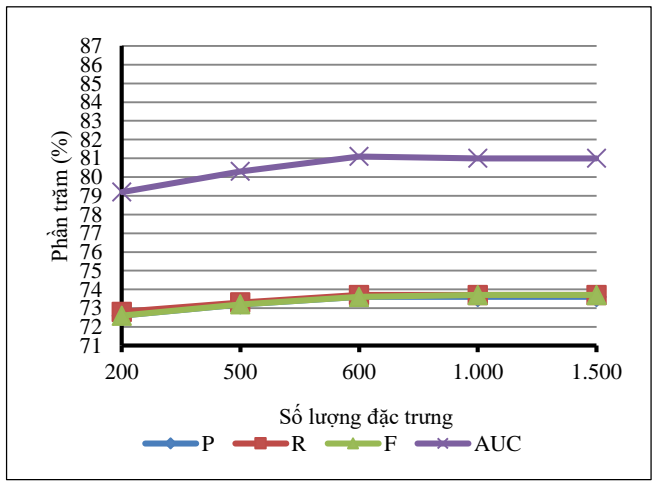

Hình 4. Kết quả phân loại của giải thuật NB

Trong bài báo, chúng tôi có giới thiệu hai cách chọn đặc trưng có khả năng phân loại cao, và Bảng 3 là kết quả phân loại đạt được dựa trên cách tính điểm Chi-Square và Odd-Ratio. Với 500 đặc trưng, việc chọn theo Chi-Square có hiệu suất tốt hơn ở cả ba giải thuật (giải thuật SVM vẫn cho kết quả tốt nhất). Tuy nhiên, sự chênh lệch của từng cách tính điểm (cùng loại giải thuật) là không đáng kể.

Các kết quả thu được ở trên, chúng tôi chỉ sử dụng tập đặc trưng gồm các từ được tách từ tập dữ liệu. Trong Bảng 4, chúng ta thấy kết quả của việc phân loại tăng lên khi thêm đặc trưng là các từ được liệt kê bằng tay. Vì vậy, từ đại diện cho các lớp chứa trong đánh giá là đặc tính quan trọng để phân loại đánh giá.

Qua các kết quả thí nghiệm, chúng tôi đã chọn phương pháp tốt nhất: giải thuật SVM với 500 đặc trưng (điểm số Chi-Square) kết hợp với các từ được liệt kê bằng tay để áp dụng lên tập kiểm thử được xử lý khác nhau: không chuẩn hóa, chuẩn hóa bằng tay, chuẩn hóa bằng phần mềm (phần mềm được phát triển bởi công ty VNG để đưa văn bản tiếng Việt về dạng chuẩn). Bảng 5 là kết quả đạt được.

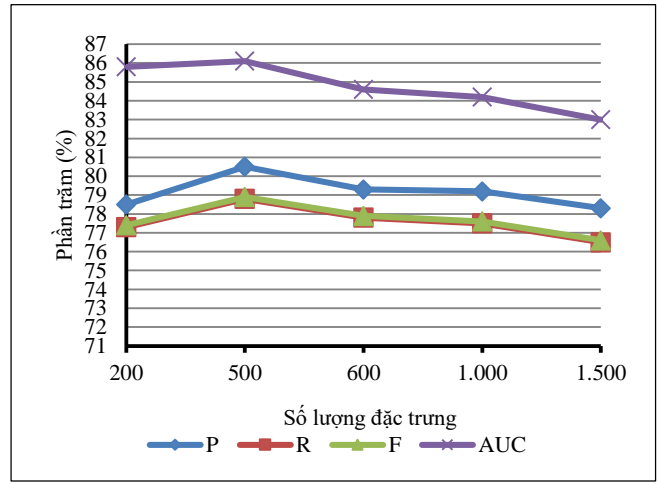

Hình 5. Kết quả phân loại của giải thuật LR

Bảng 3. Kết quả khi sử dụng hai phương pháp tính điểm

| Mô hình | Độ P | Độ R | Độ F | Độ AUC |
|---|---|---|---|---|
| 1,2$_{SVM}$ | **82,4** | **82,4** | **82,4** | 82,0 |
| 1,2$_{LR}$ | 80,5 | 78,8 | 78,9 | **86,1** |
| 1,2$_{NB}$ | 73,2 | 73,3 | 73,2 | 80,3 |
| 1,3$_{SVM}$ | 81,7 | 81,3 | 81,4 | 81,4 |
| 1,3$_{LR}$ | 79,3 | 77,6 | 77,7 | 84,7 |
| 1,3$_{NB}$ | 74,6 | 74,8 | 74,7 | 80,0 |

Bảng 4. Kết quả khi sử dụng hai loại tập đặc trưng

| Mô hình | Độ P | Độ R | Độ F | Độ AUC |
|---|---|---|---|---|
| 1,2$_{SVM}$ | 82,4 | 82,4 | 82,4 | 82,0 |
| 1,2$_{LR}$ | 80,5 | 78,8 | 78,9 | 86,1 |
| 1,2$_{NB}$ | 73,2 | 73,3 | 73,2 | 80,3 |
| 1,2,4$_{SVM}$ | **87,6** | **87,4** | **87,3** | 86,3 |
| 1,2,4$_{LR}$ | 81,4 | 80,7 | 80,8 | **88,1** |
| 1,2,4$_{NB}$ | 74,4 | 74,5 | 74,5 | 81,1 |

Bảng 5. Kết quả trên các tập kiểm thử được chuẩn hóa bằng các phương pháp khác nhau

| Dữ liệu | Độ P | Độ R | Độ F | Độ AUC |
|---|---|---|---|---|
| Không chuẩn hóa | 81,5 | 80,6 | 80,0 | 78,5 |
| Chuẩn hóa bằng phần mềm | 85,3 | 85,0 | 84,8 | 83,7 |
| Chuẩn hóa bằng tay | **87,6** | **87,4** | **87,3** | **86,3** |

## KẾT LUẬN

Trong bài báo này chúng tôi đã khảo sát và đóng góp tập dữ liệu với khoảng 50.000 đánh giá tiếng Việt. Bên cạnh đó chúng tôi đã đề xuất phương pháp sử dụng học máy để nhận biết các đánh giá spam tiếng Việt. Tập đặc trưng được chúng tôi sử dụng là sự kết hợp giữa từ được tách từ tập dữ liệu và từ được liệt kê bằng tay. Đây là tập đặc trưng lần đầu tiên được sử dụng cho bài toán phân loại đánh giá tiếng Việt. Phương pháp có kết quả nhận dạng khá tốt, cụ thể SVM cho kết quả gần 90% độ chính xác. Tuy nhiên, vì những đánh giá không thực khó phát hiện và dữ liệu về chúng còn ít nên việc tinh chỉnh phương pháp để có thể nhận dạng đánh giá không thực sẽ là hướng đi trong tương lai của chúng tôi.

## LỜI CẢM ƠN

## TÀI LIỆU THAM KHẢO